\documentclass[apjl,a4paper,12pt,useAMS]{emulateapj}
\usepackage{txfonts}
\usepackage{amssymb}

\usepackage{graphicx,graphics}
\usepackage{natbib}

\begin{document}
\title{How can newly born rapidly rotating neutron stars become magnetars?}
\author{Quan Cheng\altaffilmark{}}
\author{Yun-Wei Yu\altaffilmark{}}

\altaffiltext{}{Institute of Astrophysics, Central China Normal
University, Wuhan 430079, China, {yuyw@mail.ccnu.edu.cn}}

\begin{abstract}
In a newly born (high-temperature and Keplerian rotating) neutron
star, $r$-mode instability can lead to stellar differential
rotation, which winds the seed poloidal magnetic field ($\sim
10^{11}$ G) to generate an ultra-high ($\sim 10^{17}$ G) toroidal
field component. Subsequently, by succumbing to the Tayler
instability, the toroidal field could be partially transformed into
a new poloidal field. Through such dynamo processes, the newly born
neutron star with sufficiently rapid rotation could become a
magnetar on a timescale of $\sim 10^{2-3}$ s, with a surface dipolar
magnetic field of $\sim10^{15}$ G. Accompanying the field
amplification, the star could spin down to a period of $\sim5$ ms
through gravitational wave radiation due to the $r$-mode instability
and, in particular, the non-axisymmetric stellar deformation caused
by the toroidal field. This scenario provides a possible explanation
for why the remnant neutron stars formed in gamma-ray bursts and
superluminous supernovae could be millisecond magnetars.
\end{abstract}


\slugcomment{2014, ApJL, ??, ??}

\section{Introduction}

Since the operation of the \textit{Swift} satellite, it has been
widely suggested that the remnant compact objects formed in some
gamma-ray bursts (GRBs) could be rapidly rotating, highly magnetized
neutron stars (NSs). Such a millisecond magnetar scenario is helpful
for understanding observations such as X-ray shallow decay and, in
particular, plateau afterglows on timescales of $\sim 10^{2-4}$ s
(e.g., Dai \& Lu 1998; Zhang \& M\'esz\'aros 2001; Fan \& Xu 2006;
Yu et al. 2010; Metzger et al. 2011; Zhang 2013; Rowlinson et al.
2013) and the temporally ``extended'' gamma-ray emission on
timescales of minutes of short GRBs (Gao \& Fan 2006; Metzger et al.
2008; Bucciantini et al. 2012). Recently, a similar energy source
scenario was employed to interpret the high luminosity of some
superluminous supernovae (e.g., Kasen \& Bildsten 2010) and was also
suggested to power bright mergernova emission during the merger of a
double NS system (Yu et al. 2013). In all of these cases, the high
magnetic field of the NS is required to ensure that most of the
rotational energy of the star can be released into the stellar wind
in a sufficiently short time.

Magnetars can generally be defined as special types of NSs with
surface (though sometimes only interior) magnetic fields that are as
high as $\sim 10^{14}$ G at least. The dissipation of the magnetic
fields could power some high-energy electromagnetic emission, e.g.,
the GRB X-ray flares (Dai et al. 2006), and the bursts of soft
gamma-ray repeaters (SGRs) and anomalous X-ray pulsars (AXPs;
Thompson \& Duncan 1993). The strength of the surface dipolar
magnetic fields of most Galactic SGRs and AXPs is indeed inferred to
be on the order of $10^{14-15}$ G by
$B_{p}=3.2\times10^{19}\sqrt{P\dot{P}}$, where $P$ is the pulse
period and $\dot{P}$ is the period derivative (Olausen \& Kaspi
2013). However, the recently observed SGR 0418+5729 and Swift
J1822.3--1606 both indicate much lower dipolar fields ($6\times
10^{12}$ G and $3.8\times 10^{13}$ G, respectively; Rea et al.
2013), which are typical for normal pulsars rather than magnetars.
This ``contradiction'' hints that the high magnetic fields of some
magnetars could be dominated by multipolar (e.g., toroidal)
components, which are probably hidden in the stellar interior. The
interior fields could be much stronger than those on the surface.

The most straightforward consideration one might suggest is that the
high magnetic field of a newly born magnetar may originate from the
fossil magnetic fluxes in the progenitor core via the magnetic flux
conservation, where the progenitor should be highly magnetized
(Ferrario \& Wickramasinghe 2006). However, considering the possible
existence of interior multipolar magnetic fields in SGRs/AXPs and
the extremely rapid rotation of the GRB magnetars, it is believed
that the origin of the high magnetic field is more likely to be
associated with a dynamo process located deep in the stellar
interior. Duncan \& Thompson (1992) proposed that an
$\alpha-\omegaup$ dynamo could be supported by neutrino-driven
turbulent convection and initially existing differential rotation in
newly born millisecond NSs. Alternatively, in this Letter we suggest
that the extremely rapid rotation of newly born NSs can
spontaneously initiate a dynamo process via $r$-mode instability and
the magnetic Tayler instability.

In a rotating NS, $r$-modes arise due to the action of the Coriolis
force with positive feedback (Andersson 1998; Friedman \& Morsink
1998). The growth of the $r$-modes can be suppressed by viscous
damping and, in particular, by some non-linear effects.
Specifically, by expanding the $r$-modes up to the second order of
amplitude, the differential rotation induced by the modes can be
found to determine a saturation state of the instability (S\'a \&
Tom\'e 2005, 2006; Yu et al. 2009). As a result, a toroidal magnetic
field component can be formed and amplified by winding up the seed
poloidal field (Rezzolla et al. 2000; Rezzolla et al. 2001a, 2001b).
As it increases in a stably stratified stellar interior, the
toroidal field could enter into the Tayler instability and therefore
can be partly transformed into a new poloidal component. Finally, a
stable poloidal--toroidal twisted torus configuration appears in the
stellar interior, which is connected by an enhanced dipolar field on
the stellar surface (Braithwaite \& Spruit 2004). Such a dynamo
mechanism has been previously investigated in the framework of
accreting NS binaries (Cuofano \& Drago 2010; Cuofano et al. 2012),
where the solid crust of the NSs can provide an extra effective
suppression on the $r$-mode instability. In contrast, for a newly
born NS, the crust cannot form initially due to the high stellar
temperature. Moreover, the rotation of the newly born NS could be
very close to the Keplerian limit, which is much more rapid than
what an accreting binary NS can reach (Hessels et al. 2006).

The chief purpose of this Letter is to report on the formation of
the remnant magnetars harbored in GRBs. In the Section 2, we
describe the model for the evolutions of $r$-modes, magnetic fields,
and stellar rotations. Calculated results are presented in Section
3. Conclusion and discussions are given in Section 4.

\section{An Evolutionary Model of Newly Born Neutron Stars}
\subsection{$r$-mode Evolution}
A phenomenological model for the evolution of the $l=m=2$ mode of
primary importance was first developed by Owen et al. (1998) and
subsequently improved by Ho \& Lai (2000), who demonstrated that the
$r$-mode evolution is guided by the conservation of angular
momentum. The saturation amplitude of the $r$-mode is determined by
some nonlinear effects. Specifically, the second-order solution of
the $r$-mode gives a saturation amplitude of $\alpha_{\rm
sat}\sim(\delta+2)^{-1/2}$ (S\'a \& Tom\'e 2005), where
$-5/4\leqslant \delta \lesssim 10^{13}$ is a free model parameter
representing the initial amount of differential rotation. In the
same model framework, the physical angular momentum of the $r$-mode
can be calculated by $J_r={1\over2}(4\delta+5)\alpha^2I^*\Omega$,
where $I^*=1.635\times10^{-2}MR^2$ is an effective momentum of
inertia for the mode, $\Omega$ is the angular spin frequency, and
$M$ and $R$ are the mass and radius of the NS, respectively.

The $r$-mode angular momentum can increase through a gravitational
wave (GW) radiation back-reaction and, meanwhile, be decreased by
viscous damping and by winding the seed poloidal magnetic field to
form a toroidal component. Therefore, the evolution equation for the
$r$-mode angular momentum can be written as
\begin{eqnarray}
{d J_r\over dt}&=& 2 J_r\left({1\over\tau_{\rm
g,r}}-{1\over\tau_{\rm v}}-{1\over\tau_{\rm t}}\right) \label{Jrt},
\end{eqnarray}
where the timescales $\tau_{\rm g,r}$, $\tau_{\rm v}$, and
$\tau_{\rm t}$ correspond to the GW radiation induced by the
$r$-mode, the viscous damping, and the formation of the toroidal
magnetic field, respectively. The expressions of the two former
timescales can be found in Yu et al. (2009), while the last one is
defined as $\tau_{\rm t}=2 E_r/\dot{E}_{\rm t}$, where
$E_r={1\over2}(4\delta+9)\alpha^2I^*\Omega^2$ is the energy of the
$r$-mode and $\dot{E}_{\rm t}$ is the change rate of the toroidal
fields energy. For simplicity, some magnetic back reactions are not
taken into account in Equation (\ref{Jrt}). When the toroidal fields
energy becomes comparable to the kinetic energy of the differential
rotation, the Lorentz force exerted on the plasma would approach to
reverse the differential rotation (Braithwaite 2006a). Then a
dynamic equilibrium could be built at which point the toroidal field
formation can no longer be regarded as a dissipation process. Such
an effect could suppress the peak strength of the magnetic fields
that are presented in Section 3 by a factor of a few. In more
detail, a sufficiently strong Lorentz force could affect the drift
velocity of the given fluid element, the azimuthal displacement, and
the rate of energy transfer (e.g., Morsink \& Rezania 2002).

\subsection{Magnetic Field evolution}
Secular azimuthal drifts on the isobaric surfaces due to the
differential rotation gradually generate a large scale azimuthal
magnetic field. Following Rezzolla et al. (2000), S\'a (2004), S\'a
\& Tom\'e (2005), and Cuofano \& Drago (2010), the strength of the
azimuthal field at coordinate, $\textbf{r}$, in the star at a given
time, $t$, can be calculated by
\begin{eqnarray}
B^{\phi}(\textbf{r},t)=B_{\rm d}\left({R\over
r}\right)^3(4\cos^2\theta+\sin^2\theta)|\xi^{\phi}(\textbf{r},t)|,
\end{eqnarray}
where an internal dipolar magnetic field with a surface strength,
$B_{\rm d}$, is assumed and
\begin{eqnarray}
\xi^{\phi}(\textbf{r},t)={15\over 32\pi}\left({r\over
R}\right)^2\sin^2\theta (2\delta+3)\int_0^t\alpha^2\Omega dt'
\end{eqnarray}
is the total azimuthal displacement from the onset of the $r$-mode
instability to time, $t$. The increase rate of the total energy of
the toroidal field can be calculated by integrating over the whole
stellar volume
\begin{eqnarray}
{dE_{\rm t}\over dt}&=&{d\over
dt}\left(\int{{{B^{\phi}}^{2}(\textbf{r},t)\over
8\pi}dV}\right)\nonumber\\&\approx&{15\over 56\pi^2}B_{\rm
d}^2R^3(2\delta+3)^2\alpha^2\Omega\int_0^t\alpha^2\Omega dt'.
\end{eqnarray}
Here, we further define a volume-averaged strength of the toroidal
field by using $E_{\rm t}=(4\pi R^3/3)(\bar{B}^2_{\rm t}/8\pi)$, and
then the evolution of $\bar{B}_t$ can be determined by
\begin{eqnarray}
{d\bar{B}_{\rm t}\over dt}\approx {3\over2\pi}\left({5\over
14}\right)^{1/2}B_{\rm d}(2\delta+3){\alpha^2\Omega}.\label{Bphit}
\end{eqnarray}
A quadrupolar deformation of the NS may appear with an ellipticity
of $\epsilon=-{{5\bar{B}_{\rm t}^2R^4}/{6GM^2}}$, an important
consequence of the toroidal fields formation. Such a deformation
will cause the NS to produce additional GW radiation that is much
stronger than that induced by the $r$-mode itself.

It is further expected that the ultra-high toroidal field will
succumb to the Tayler instability that closes the dynamo loop by
generating a new poloidal field. Here, it is required that the spin
frequency is lower than the Alfv\'en frequency $\omega_{\rm
A}=\bar{B}_{\rm t}/\left(R\sqrt{4\pi \bar{\rho}}\right)$ of the
stellar material (Braithwaite 2006b). In other words, the timescale
of the Tayler instability, $\tau_{\rm TI}\sim 2\pi/\omega_{\rm A}=
2.2\ {\rm ms}\left(\bar{B}_{\rm t}/10^{17}{\rm G}\right)^{-1}$,
should be shorter than the spin period of the star. Such a condition
could be satisfied in a short time due to the increase of
$\bar{B_{\rm t}}$ and the decrease of $\Omega$ by GW radiation. As a
result, a poloidal--toroidal twisted torus shape can be built in the
stellar interior. Nevertheless, some previous studies for a stable
magnetic configuration suggested that the poloidal component is
probably overwhelmingly subordinate to the toroidal one (e.g.,
Mastrano et al. 2011). Moreover, since some poloidal field lines
could be closed in the stellar interior, the field that extends to
the stellar surface to connect with the outer dipolar field could be
much weaker than the internal toroidal one. Therefore, in the
following calculations, we will adopt
\begin{eqnarray}
B_{\rm d}=\xi\bar{B}_{\rm t}\label{Bp} ~~{\rm
for}~~\Omega<\omega_{\rm A}
\end{eqnarray}
with a reference value of $\xi =0.01$. This assumption could be
supported by the fact that the surface dipolar magnetic field of
some SGRs is inferred to be lower than the internal field (Stella et
al. 2005; Dall'Osso et al. 2009).

\subsection{Spin Evolution}
A rapidly rotating, newly born NS could be spun down by GW radiation
and magnetic dipole radiation, whereas the former can be due to both
the $r$-mode oscillation and the magnetic deformation of the star.
Therefore, the decrease of the total stellar angular momentum can be
written as
\begin{equation}
{dJ\over dt}= -{3\alpha^2I^*\Omega\over \tau_{\rm
g,r}}-{I\Omega\over \tau_{\rm g,t}}-{I\Omega\over \tau_{\rm
d}}\label{Jt},
\end{equation}
where $J=I\Omega+J_r$ with $I=0.261MR^2$ being the stars moment of
inertia. The timescales corresponding to the magnetic dipole
radiation read $\tau_{\rm d}={6Ic^3/( B_{\rm
d}^2R^6\Omega^2\sin^2\chi)}$ and to the GW radiation due to magnetic
deformation reads $\tau_{\rm
g,t}=5c^5/[2GI\epsilon^2\Omega^4\sin^2\chi(1+15\sin^2\chi)]$ (Cutler
\& Jones 2001), where $\chi$ is the inclined angle between the
magnetic and spin axes. The initial value of $\chi$ could be close
to zero, but a deviation between the two axes is also expected to
happen quickly though the details of the processes are uncertain
(e.g., Dall'Osso et al. 2009; Cutler 2002).

Combining Equations. (\ref{Jrt}) and (\ref{Jt}), we can obtain the
coupled evolution equations for the $r$-mode amplitude and spin
frequency as
\begin{eqnarray}
{d\alpha\over dt}&=&\left[1+{2\alpha^2\over
15}(\delta+2)\right]{\alpha\over \tau_{\rm
g,r}}\nonumber\\
&-&\left[1+{\alpha^2\over 30}(4\delta+5)\right]\left({\alpha\over
\tau_{\rm v}}+{\alpha\over \tau_{\rm t}}\right)+{\alpha\over
2\tau_{\rm g,t}}+{\alpha\over 2\tau_{\rm d}},\label{alpha}
\\
{d\Omega\over dt}&=&-{4\alpha^2\over 15}(\delta+2){\Omega\over
\tau}-{\Omega\over \tau_{\rm g,t}}-{\Omega\over \tau_{\rm
d}},\label{Omega}
\end{eqnarray}
where $I^*/I\approx 1/15$ is taken and the timescale, $\tau$, reads
\begin{eqnarray}
\tau=\left[{1\over \tau_{\rm
g,r}}-{(4\delta+5)\over4(\delta+2)}\left({1\over\tau_{\rm
v}}+{1\over\tau_{\rm t}}\right)\right]^{-1}.
\end{eqnarray}

\section{Results}\label{Sec III}
From Equations. (\ref{Bphit}), (\ref{Bp}), (\ref{alpha}), and
(\ref{Omega}), we can calculate the strengths of the magnetic
fields, the spin frequency, and the $r$-mode amplitude as functions
of time since the birth of an NS, where an analytical cooling
history dominated by a modified Urca process is adopted as $T=T_{\rm
i}(1+t/\tau_{\rm c})^{-1/6}$ with $\tau_{\rm c}=20(T_{\rm
i}/10^{10}\rm K)^{-6}$ s and $T$ represents the stellar temperature.
Our results reveal that the secular evolutions of these quantities
are very insensitive to the initial values of $T_{\rm i}$,
$\alpha_{\rm i}$, and $B_{\rm d,i}$ within a wide parameter range,
whereas the initial spin frequency could significantly influence the
evolutions. By taking $T_{\rm i}=10^{10}$ K, $\alpha_{\rm
i}=10^{-10}$, and $B_{\rm d,i}=10^{11}$ G, which is typical for
normal pulsars, we plot the evolution curves of $\bar{B}_{\rm t}$,
$B_{\rm d}$, $\nu$ ($=\Omega/2\pi$), and $\alpha$ in Figure 1 for
three different initial spin frequencies as $\Omega_{\rm
i}=\Omega_{\rm K}$, $(3/4)\Omega_{\rm K}$, and $(1/2)\Omega_{\rm
K}$, where $\Omega_{\rm K}=2\sqrt{\pi G\bar{\rho}}/3$ is the
Keplerian spin frequency.
\begin{figure}
\resizebox{\hsize}{!}{\includegraphics{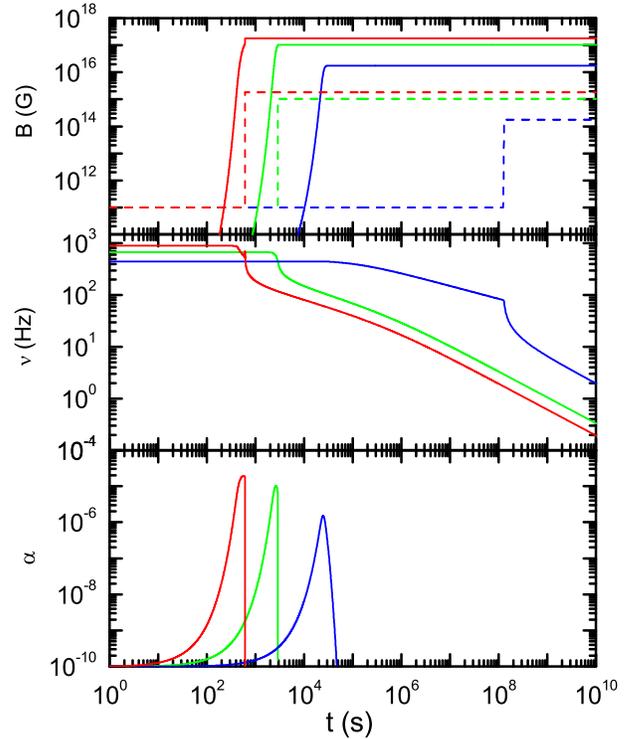}}
\caption{Evolutions of the strengths of the magnetic fields (top),
spin frequency (middle), and $r$-mode amplitude (bottom). Red,
green, and blue lines correspond to an initial angular spin
frequency of $\Omega_{\rm K}$, $(3/4)\Omega_{\rm K}$,
and$(1/2)\Omega_{\rm K}$, respectively. The solid and dashed lines
in the top panel correspond to the internal toroidal and surface
dipolar magnetic fields, respectively.} \label{Fig1}
\end{figure}

The solid red and green lines in the top panel of Figure 1 show
that, for $\Omega_{\rm i}\geqslant(3/4)\Omega_{\rm K}$, the toroidal
magnetic field can be increased to as high as a few times $10^{17}$
G on a timescale of $\sim10^{2-3}$ s. Subsequently, the high
quadrupolar ellipticity $\epsilon\sim 0.01$ of the NS gives rise to
a strong GW radiation, which leads the NS to spin down with a
temporal behavior of $\propto t^{-1/4}$, as shown in the middle
panel of Figure 1 (for $\chi=\pi/2$). Strictly speaking, the start
time of this GW braking phase is determined by the uncertain
deviation of the magnetic axis from the spin axis. Nevertheless, for
$\Omega_{\rm i}\geqslant(3/4)\Omega_{\rm K}$, such an uncertainty
may not influence the stellar magnetic evolution, because the
ultra-high toroidal field can make the timescale of the Tayler
instability shorter than the spin period even though the NS is only
slightly spun down by the $r$-mode-induced GW radiation. So, the
surface dipolar magnetic field of the NS can be amplified to $\sim
10^{15}$ G on the same timescale of $\sim10^{2-3}$ s. However, for
lower initial spin frequencies (e.g., $\Omega_{\rm
i}=(1/2)\Omega_{\rm K}$; the blue lines in Figure 1), the consequent
lower toroidal field ($\sim10^{16}$ G) would determine a Tayler
instability timescale much longer than the initial spin period
($P_{\rm i}=2\pi/\Omega_{\rm i}$). Hence, the occurrence of the
Tayler instability requires a remarkable spin-down of the NS, which
is beyond the ability of the $r$-mode-induced GW radiation.
Therefore, a long-time braking by the GW radiation, due to the
magnetic deformation, becomes necessary. The amplification of the
surface field is delayed until it is too late (e.g., $\sim 10^8$
s)\footnote{On such a long timescale, more complexity will be
involved. For $t\gtrsim 10^7$ s, the temperature of the NS dropped
to $T\lesssim10^9$ K, at which a stellar crust and a superconducting
layer could form. Therefore, if the field amplification happens
later than $10^7$ s, the emergence of the amplified field from the
stellar surface will be seriously suppressed by the solid crust for
an extremely long time.} to be consistent with the GRB timescales.
The magnetar's formation should occur much earlier than the
GRB-associated supernova on a timescale of $\sim10^6$ s. More
calculations will reveal a critical initial spin period of $P_{\rm
i,c}=1.7$ ms. When $P_{\rm i}\leqslant P_{\rm i,c}$, the Tayler
instability can happen with only $r$-mode-induced GW radiation, and
the surface field can be simultaneously amplified to the generation
of the toroidal field.

\begin{figure}
\resizebox{\hsize}{!}{\includegraphics{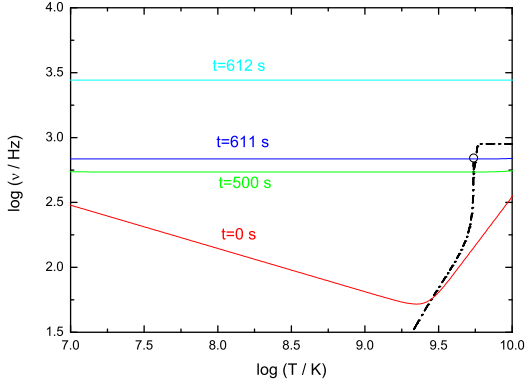}} \caption{$R$-mode
instability windows (regions above the solid lines) at different
times, as labeled, which evolve with time because of the magnetic
evolution. The dash--dotted line represents the evolution trajectory
of an NS for $B_{\rm d,i}=10^{11}$ G, $\Omega_{\rm i}=\Omega_{\rm
K}$, and $T_{\rm i}=10^{10}$ K in the $\nu--T$ plane. The open
circle on the evolution curve indicates the time of 611 s at which
the evolution curve crosses the boundary of the instability window.}
\label{Fig2}
\end{figure}

For the $r$-mode evolution, the bottom panel of Figure 1 shows that
the maximum amplitude of the $r$-mode is restricted to $\lesssim
10^{-5}$ for the adopted parameter $\delta=10^{10}$, which is taken
to be consistent with the saturation amplitude determined by some
other possible nonlinear effects. For example, Bondarescu et al.
(2007) revealed a saturation amplitude on the order of $\sim
10^{-5}$ by coupling the $r$-modes with other two inertial modes. In
fact, a higher saturation amplitude would not significantly change
the toroidal magnetic fields because the $r$-mode energy,
$E_{r}\propto\alpha_{\rm sat}^2\delta$, is weakly dependent on the
parameter. On the other hand, the duration of the instability is
restricted to $\sim 10^{2-4}$ s for $\Omega_{\rm i}\geqslant
(1/2)\Omega_{\rm K}$, which is drastically shorter than that
obtained without the consideration of the magnetic field evolution
(Yu et al. 2009). For a more general understanding, in Figure 2, we
display the temporal-dependent $r$-mode instability windows, the
boundaries of which (solid lines) are determined by the equation
$dJ_{r}/dt=0$. As shown, the instability window shrinks to the
high-$\nu$ region very quickly and becomes temperature-independent.
Such a window evolution is caused by the magnetic field evolution,
because the dissipation of the $r$-mode is primarily through the
energy transfer from the $r$-mode to the toroidal field. As a
result, the $r$-mode instability could be switched off at very early
time, even though the star has only slightly spun down. The stellar
temperature at which the $r$-mode instability ends can be found to
be around $5.5\times10^{9}$ K. Therefore, the solid crust of the NS
could not have been formed during the action of the $r$-mode
instability (Chamel \& Haensel 2008; Dall'Osso et al. 2009),
therefore, the damping effects arising from the boundary of the
crust (Mendell 2001) can be ignored. Finally, the instability window
for $t=0$ s shows that, for an initial spin period longer than
$\sim$3 ms, the $r$-mode instability can be effectively suppressed
by the viscous damping.

\section{CONCLUSION AND DISCUSSIONS}\label{Sec V}
By considering the differential rotation caused by $r$-mode
instability in a newly born, rapidly rotating NS, we calculate the
evolution of the stellar magnetic fields, where an ultra-strong
toroidal magnetic field is generated. Succumbing to the Tayler
instability, the toroidal field is partially transformed into a new
poloidal field. Through such dynamo processes, the NS could become a
magnetar with a surface dipolar field of a strength $\sim 10^{15}$ G
on timescales $\sim 10^{2-3}$ s, the precondition of which is that
the NS should rotate initially with a nearly Keplerian period,
$P_{\rm i}\lesssim 1.7$ ms. Such a condition could easily be
satisfied in the situation of GRBs. For somewhat longer periods,
$1.7{\rm ms}<P_{\rm i}\lesssim3$ ms, this dynamo could work in
principle, but the strengths of the fields become much lower.
Moreover, the amplification of the surface field is delayed to a
very late time, at which more complexity (e.g., the formation of a
crust) is involved. In any case, the long time delay could make the
model inapplicable for GRB magnetars. Finally, for $P_{\rm i}>3$ ms,
the dynamo processes would never happen and a normal magnetic field
keeps in the NS, because the $r$-mode instability is suppressed by
viscosities.

Due to the magnetic dissipation, the $r$-mode-induced GW radiation
becomes very weak. Alternatively, another strong GW radiation is
produced due to the high deformation of the NS by the toroidal
magnetic field, which could cause the star to be a promising target
for GW detection. As a result, accompanying the magnetic field
amplification, the spin periods of GRB magnetars would be increased
to $\sim 5$ ms. In other words, the ``initial" spin periods derived
from GRB afterglow observations should be basically consistent with
such a value. Furthermore, due to the GW radiation, a remarkable
amount of the rotational energy of the NS can be released into the
GW. Therefore, the supernova remnant around the magnetar cannot be
as highly energized as usually considered. In observation, analysis
of the X-ray spectra of some supernova remnants associated with
magnetar candidates Vink \& Kuiper (2006) revealed that the total
energy in these supernova remnants is almost nothing, which may
favor our model. In other words, some Galactic magnetars may share
the same origin mechanism presented here.

\acknowledgements This work was motivated by the discussions on
magnetar GW radiation with R. X. Xu when Y. -W. Y. visited Peking
University in the winter of 2012. The authors thank the anonymous
referee, X. P. Zheng, B. Zhang, and H. Yang for their helpful
comments. This work is supported by the 973 program (grant No.
2014CB845800), the National Natural Science Foundation of China
(grant Nos. 11103004, 11073008, and 11178001), the funding for the
Authors of National Excellent Doctoral Dissertations of China (grant
No. 201225), and the Program for New Century Excellent Talents in
University (grant No. NCET-13-0822).


\begin{thebibliography}{99}

\bibitem{Andersson:1998}Andersson, N. 1998, ApJ, 502, 708

\bibitem{Bondarescu:2007}Bondarescu, R., Teukolsky, S. A., Wasserman, I. 2007, Phys. Rev. D,
76, 064019

\bibitem{Braithwaite:2006a}Braithwaite, J. 2006a, Astron. Astrophys., 449, 451

\bibitem{Braithwaite:2006b}Braithwaite, J. 2006b, Astron. Astrophys., 453, 687

\bibitem{Braithwaite:2004}Braithwaite, J., Spruit, H. C. 2004, Nature, 431, 819

\bibitem{Bucciantini:2012}Bucciantini, N., Metzger, B. D., Thompson, T. A., Quataert, E. 2012, MNRAS, 419, 1537

\bibitem{Chamel:2008}Chamel, N., Haensel, P. 2008, Living Rev. Relativity, 11, 10

\bibitem{Cuofano:2012}Cuofano, C., Dall'Osso, S., Drago, A., and Stella, L. 2012, Phys.Rev. D, 86, 044004

\bibitem{Cuofano:2010}Cuofano, C., Drago, A. 2010, Phys.Rev. D, 82, 084027

\bibitem{}Cutler, C. 2002, Phys. Rev. D, 66, 084025

\bibitem{Cutler:2001}Cutler, C., Jones, D. I. 2001, Phys. Rev. D, 63, 024002

\bibitem{Dai:1998}Dai, Z. G., Lu, T. 1998, Phys. Rev. Lett., 81, 4301

\bibitem{Dai:2006}Dai, Z. G., Wang, X. Y., Wu, X. F., Zhang, B. 2006, Science, 311, 1127

\bibitem{Dall:2009}Dall'Osso, S., Shore, S. N., Stella, L., 2009, MNRAS, 398, 1869

\bibitem{Duncan:1992}Duncan, R. C., Thompson, C. 1992, ApJL, 392, L9

\bibitem{Fan:2006}Fan, Y. Z., \& Xu, D. 2006, MNRAS, 372, L19

\bibitem{Ferrario:2006}Ferrario, L., Wickramasinghe, D. 2006, MNRAS, 367, 1323

\bibitem{Friedman:1998}Friedman, J. L., Morsink, S. M. 1998, ApJ, 502, 714

\bibitem{Gao:2006}Gao, W. H. \& Fan, Y. Z. 2006, ChJAA, 6, 513

\bibitem{Gompertz:2013}Gompertz, B. P., O'Brien, P. T., Wynn, G. A., Rowlinson,
A. 2013, MNRAS, 431, 1745

\bibitem{Hessels:2006}Hessels, J. W. T., Ransom, S. M., Stairs, I. H., Freire, P. C. C.,
Kaspi, V. M., Camilo, F. 2006, Science, 311, 1901

\bibitem{Ho:2000}Ho, W. C. G., Lai, D. 2000, ApJ, 543, 386

\bibitem{Kasen:2010}Kasen, D., Bildsten, L. 2010, ApJ, 717, 245

\bibitem{Mastrano:2011}Mastrano, A., Melatos, A., Reisenegger, A., Akg\"un, T. 2011, MNRAS, 417, 2288

\bibitem{Mendell:2001}Mendell, G. 2001, Phys.Rev. D, 64, 044009

\bibitem{Metzger:2008}Metzger, B. D., Quataert, E., Thompson, T. A. 2008, MNRAS, 385, 1455

\bibitem{Morsink:2002}Morsink, S. M., Rezania, V. 2002, ApJ, 574, 908

\bibitem{Olausen:2013}Olausen, S. A., Kaspi, V. M. 2013, arxiv: 1309.4167

\bibitem{Owen:1998}Owen, B. J., Lindblom, L., Cutler, C., Schutz, B. F.,
Vecchio, A., Andersson, N. 1998, Phys. Rev. D, 58, 084020

\bibitem{Rea:2013}Rea, N. et al. 2013, ApJ, 770, 65

\bibitem{Rezzolla:2001a}Rezzolla, L., Lamb, F. K., Markovi$\rm \acute{c}$, D., Shapiro, S. L.
2001a, Phys. Rev. D., 64, 104013

\bibitem{Rezzolla:2001b}Rezzolla, L., Lamb, F. K.,
Markovi$\rm \acute{c}$, D., Shapiro, S. L. 2001b, Phys. Rev. D., 64,
104014

\bibitem{Rezzolla:2000}Rezzolla, L., Lamb, F. K., Shapiro, S. L. 2000, ApJL, 531, L141

\bibitem{Rowlinson:2013}Rowlinson, A. et al. 2013, MNRAS, 430, 1061

\bibitem{Sa:2004}S$\rm \acute{a}$, P. M. 2004, Phys. Rev. D, 69, 084001

\bibitem{Sa:2005}S\'a, P. M., Tom\'e, B. 2005, Phys. Rev. D, 71, 044007

\bibitem{Sa:2006}S\'a, P. M., Tom\'e, B. 2006, Phys. Rev. D, 74, 044011

\bibitem{Stella:2005}Stella, L., Dall'Osso, S., Israel, G. L., Vecchio, A. 2005, ApJL, 634,
L165

\bibitem{Thompson:1993}Thompson, C., Duncan, R. C. 1993, ApJ, 408, 194

\bibitem{Vink:2006}Vink, J., Kuiper, L. 2006, MNRAS, 370, L14

\bibitem{Yu:2009}Yu, Y. W., Cao, X. F., Zheng, X. P. 2009, Research in Astron. and
Astophys., 9, 1024

\bibitem{Yu:2010}Yu, Y. W., Cheng, K. S., Cao, X. F. 2010, ApJ, 715, 477

\bibitem{Yu:2013}Yu, Y. W., Zhang, B., Gao, H. 2013, ApJL, 776, L40

\bibitem{Zhang:2013}Zhang, B. 2013, ApJL, 763, L22

\bibitem{Zhang:2001}Zhang, B., M\'esz\'aros, P. 2001, ApJL, 552, L35

\end{thebibliography}
\end{document}